\documentclass[showpacs]{revtex4-1}
\usepackage{graphicx}
\usepackage{dcolumn}
\usepackage{bm}
\usepackage{bm}
\usepackage{amsmath}
\usepackage[percent]{overpic}
\usepackage{subfig}
\usepackage{amssymb}
\begin{document}

\title{Coherent-State Overcompleteness, Path Integrals, and Weak Values}
\author{Fernando Parisio}
\email{parisio@df.ufpe.br} 
\address{Departamento de F\'{\i}sica, Universidade Federal de Pernambuco, 50670-901,
Recife, Pernambuco, Brazil}

\begin{abstract}
In the Hilbert space of a quantum particle the standard coherent-state resolution of unity is written in terms of a phase-space integration of the outer product $|z\rangle \langle z|$. Because no pair of coherent states is orthogonal, one can represent the closure relation in non-standard ways, in terms of a single phase-space integration of the ``unlike'' outer product $|z'\rangle \langle z|$, $z'\ne z$. We show that all known representations of this kind have a common ground, and that our reasoning extends to spin coherent states.
These unlike identities make it possible to write formal expressions for a phase-space path integral, where the role of the Hamiltonian ${\cal H}$ is played by a weak energy value ${\cal H}_{weak}$. Therefore, in this context, we can speak of weak values without any mention to measurements. The quantity ${\cal H}_{weak}$ appears as the ruler of the phase-space dynamics in the semiclassical limit. 
\end{abstract}

\maketitle


\section{Introduction: Single basis, different representations}
\label{intro}
In quantum mechanics the term overcompleteness is used to designate a redundant set of vectors that spans a system's 
Hilbert space. We loosely refer to such a set as an overcomplete {\it basis}, while it would be preferable to use the term {\it  tight frame} \cite{ole}. Since the former is a widely used terminology we will employ it, keeping in mind that it is not a minimal generating set.
The use of a frame of this type gives rise to an infinity of representations in the sense that a
single ket can be decomposed in different ways in terms of the same set of vectors.
In this regard the terms ``choosing a basis" and ``choosing a representation", which are routinely used
interchangeably, are no longer equivalent.

Why would one bother to represent a vector in a potentially ambiguous way?
It turns out that the elements of certain overcomplete basis have a relevant physical meaning
and mathematical properties that have proved to be very helpful.
The use of canonical coherent states $\{| z \rangle \}$, arguably the most important overcomplete 
set in physics \cite{glauber,glauber2}, is more than sufficient to illustrate the physical relevance of these sets. 

For a basis to be fully operational, one should be able to write a closure relation in terms of it. 
The standard way to do that with coherent states is to represent the unit operator 
by
\begin{equation}
\hat{I} = \int \frac{{\rm d}^2 z}{\pi} | z \rangle \langle z| \;,
\label{I0}
\end{equation}
where $z$ is a complex label and the integration is over its real and imaginary parts.

The overcompleteness of $\{ |z \rangle\}$ is a direct consequence of 
the analyticity of the Bargmann function $\psi(z^*)=\exp\{+|z|^2/2\}\langle z|\psi\rangle$ \cite{bargmann}. 
For example, 
$\{| z_j \rangle \}$ with $\{z_j\}_{j \in \mathbb{N}} $ being a convergent sequence on the complex plane, has
been shown to constitute a basis \cite{cahill}. As a corollary, it is possible to represent any ket in terms of coherent states belonging to any curve 
with non-zero length in the complex plane.
Works can be found in the literature in which
different subsets of $\{| z \rangle \}$ are used to construct alternative representations. 
An interesting example is the circle decomposition, in which an arbitrary ket can be written as
\begin{equation}
| \psi \rangle= \frac{e^{R^2/2}}{2 \pi i}\oint_{|z|=R} {\rm d} z\, g(z) | z \rangle \;,
\label{circle}
\end{equation}
where only coherent states on the circle of radius $R$ are used \cite{circle}.
In a related representation, only coherent states of vanishing momentum are employed
(corresponding to the real axis in the $z$-plane) \cite{line,line2}.
These and other \cite{szabo} examples make it clear the redundant character of $\{ |z\rangle\}$.

In addition, there is another property that is particularly relevant: two arbitrary coherent states are never orthogonal.
This property has an immediate consequence, namely, the unit operator can also 
be expressed as 
\begin{equation}
\hat{I} = \hat{I}^2= \int \int \frac{{\rm d}^2 z}{\pi} \frac{{\rm d}^2 z'}{\pi}| 
z \rangle \langle z| z' \rangle \langle z'| \;,
\label{I0b}
\end{equation}
which is not trivially equivalent to (\ref{I0}).
Note that in the $\{| x \rangle \}$ representation, e.g, the analogous of (\ref{I0b}) would be identical to 
that corresponding to (\ref{I0}), since $\langle x|x'\rangle=\delta(x-x')$. The significance of relations 
(\ref{I0}) and (\ref{I0b}) is described by Klauder and Sudarshan as ``two manifestly different decompositions 
for the same operator in terms of one set of states" \cite{klauder}. They also describe (\ref{I0}) as involving 
a superposition of ``like outer products" in opposition to (\ref{I0b}) which is a composition of ``unlike outer products"
\cite{klauder}, involving a double phase-space integration.

Despite the many alternative representations, the only existing expressions
for the unit operator in terms of coherent states were (\ref{I0}) and its immediate consequences, e.g., (\ref{I0b}).
The only exceptions being a result by Solari \cite{solari},
and
\begin{equation}
\hat{I} = \int \frac{{\rm d}^2 z}{\pi} \lambda e^{ \frac{1}{2}(1-\lambda)^2 |z|^2} | \lambda z \rangle \langle z| \;,
\label{I1}
\end{equation}
with $\lambda$ being a positive real number,
derived in \cite{parisio}, both employing ``unlike outer products"
inside a {\it single} phase-space integration. The above identity can be pictorially understood as follows. Coherent-state overcompleteness not only imply in redundancy, but also in non-orthogonality, ($\langle z| z' \rangle \ne 0$). Therefore, we can take the component of an arbitrary ket $|\psi\rangle$, $\langle z|\psi\rangle$, and force it to be the coefficient of $|\psi\rangle$ in the ``wrong direction'' $|\lambda z\rangle$ provided that we correct this by an appropriate measure in phase space. We will return to this point shortly.

Identities like (\ref{I1}) considerably add to the multifold ambiguity in the definition of coherent-state path integrals \cite{klauder,marcus} in a way that is not directly related to ordering, but rather to the off-diagonal character of the identities.

Our objective in this work is twofold. First we provide a common ground for unlike closure relations like those in \cite{parisio,solari,solari2} also presenting a generalization of (\ref{I1}) to spin coherent states, and, second, we employ the Solari identity to develop a phase-space path integral for which the contributing trajectories obey dynamical equations with the classical Hamiltonian $\cal{H}$ replaced by a weak energy value as originally defined by Aharonov, Albert, and Vaidman \cite{aharonov}:
\begin{equation}
\label{weak}
{\cal H}_{weak}=\frac{\langle\psi_f|\hat{H}| \psi_0 \rangle}{\langle \psi_f| \psi_0 \rangle}\;,
\end{equation}
where the states $|\psi_0\rangle$ and $|\psi_f\rangle$ are such that $\langle \psi_0|\psi_f\rangle \approx 1$. The manuscript is organized as follows. In the next section we briefly list some basic properties of coherent states. In section III we re-derive the relation given in \cite{solari} in a clearer, algebraic way. We then proceed, in section IV, to derive our phase-space path integral and discuss its semiclassical limit. Finally, in section V, we call attention to some recurrences in quantum phase-space transforms, and summarize our conclusions. In the appendix we use an overcomplete set in $\mathbb{R}^2$ to make our point about unlike closure relation explicit in the simplest possible framework. 
\section{Canonical coherent states}
\label{sec:1}

There are entire books dedicated to the properties and applications of coherent states. Classical examples are those by Perelomov 
\cite{perelomov} and by Klauder and Skagerstam \cite{klauder2}. A more recent account is the book by Gazeau \cite{gazeau}. In this section we list a few properties of canonical coherent states that will be directly used in the remainder of this work.

Given an harmonic oscillator with mass $m$ and angular frequency $\omega$, a canonical coherent state $| z \rangle$ is defined by the eigenvalue equation $\hat{a}| z \rangle=z| z \rangle$, where
$\hat{a}=( \hat{q}/b+ib\hat{p}/\hbar)/\sqrt{2}$
is the bosonic annihilation operator and $b=\sqrt{\hbar/m\omega}$. An equivalent and insightful definition is $| z \rangle=\hat{D}(z)|0\rangle$, with $|0\rangle$ being the ground state of the harmonic oscillator, and the displacement operator is given by 
\begin{equation}
\hat{D}(z)=\exp\{z\hat{a}^{\dagger}-z^*\hat{a}\}=e^{-\frac{1}{2}|z|^2}e^{z\hat{a}^{\dagger}}e^{-z^*\hat{a}}\;,
\label{displace}
\end{equation}
where, in the second equality, we used the Baker-Haussdorf formula.
\section{Unlike closure relations}

\subsection{Unlike coherent-state closure relations}
We start by realizing that the unlike, or off-center, closure relation (\ref{I1}) derived in \cite{parisio} can be written in a more suggestive way as (for an elementary example see the appendix):
\begin{equation}
\hat{I} = \int \frac{\lambda\, {\rm d}^2 z}{\pi} \frac{ | \lambda z \rangle \langle z|}{ \langle z|
\lambda z\rangle } \;,
\label{I2}
\end{equation}
where $\frac{{\rm d}^2 z}{\pi}$ is replaced by $\frac{\lambda\, {\rm d}^2 z}{\pi}$, since $\lambda$ is the Jacobian determinant
associated to the linear transformation $z \rightarrow \lambda z$ and $z^* \rightarrow z^*$, assuming that $z$ and $z^*$
are independent variables. This is an usual procedure that is justified by the analytical extension of $\Re(z)\propto q$ and $\Im(z)\propto p$
into the complex plane (see \cite{marcus}). 

Next we give an alternative derivation of a result by Solari. Since it was presented as a side result in the appendix of \cite{solari}, and seems to remain largely unknown, we believe this alternative derivation is in order. We do it in terms of what we may call Weyl-like outer products. Define
\begin{equation}
\hat{\cal B}
= \int \frac{{\rm d}^2 z}{\pi} \, e^{-\frac{1}{2}|\zeta|^2-\zeta^*z+\zeta z^*}\,| z-\zeta \rangle \langle z+\zeta| \;,
\label{B}
\end{equation}
where $\zeta$ is an arbitrary complex number, representing a point in phase space. 

We start by disassembling (\ref{B}) in terms of its constituent displacement operators [see (\ref{displace})]. 
Note that
\begin{equation}
\hat{D}(z+\zeta)
=\exp\left\{-\frac{1}{2}\zeta z^*+\frac{1}{2}\zeta^* z \right\}\hat{D}(\zeta)\hat{D}(z)\;,
\end{equation}
which readily leads to
$|z-\zeta \rangle \langle z+\zeta|=\exp\left\{+\zeta z^*-\zeta^* z \right\}\hat{D}(-\zeta)|z\rangle \langle z|\hat{D}^{\dagger}(\zeta)$.
Replacing this relation in (\ref{B}) we obtain
\begin{eqnarray}
\nonumber
\hat{\cal B} = \hat{D}(-\zeta) \int \frac{{\rm d}^2 z}{\pi} \frac{e^{-\zeta^*z} | z \rangle \langle z|e^{\zeta z^*}}{\langle z+\zeta|z-\zeta\rangle} \hat{D}^{\dagger}(\zeta) \\
\nonumber
= e^{2|\zeta|^2}\hat{D}(-\zeta) \int \frac{{\rm d}^2 z}{\pi} \,e^{-2\zeta^*z} | z \rangle \langle z|e^{2\zeta z^*} \hat{D}^{\dagger}(\zeta) \\
\nonumber
= e^{2|\zeta|^2}\hat{D}(-\zeta) e^{-2\zeta^*\hat{a}}\left\{\int \frac{{\rm d}^2 z}{\pi} \, | z \rangle \langle z|\right\}e^{2\zeta \hat{a}^{\dagger}} \hat{D}^{\dagger}(\zeta)\\
=e^{2|\zeta|^2}\hat{D}(-\zeta)e^{-2\zeta^*\hat{a}}e^{2\zeta \hat{a}^{\dagger}} \hat{D}^{\dagger}(\zeta)=\hat{D}(-\zeta)\hat{D}(2\zeta)\hat{D}(-\zeta)=\hat{I}\;,
\end{eqnarray}
where we used $\hat{D}^{\dagger}(\zeta)=\hat{D}(-\zeta)$ and the Baker-Hausdorff formula.
Thus we prove that (\ref{B}) is a genuine resolution of unit and, in addition, we note that it can also be written in the form,
\begin{equation}
\hat{I}=\int \frac{{\rm d}^2 z}{\pi} \frac{ | z-\zeta \rangle \langle z+\zeta|}{ \langle z+\zeta|z-\zeta\rangle }\;,
\label{Iweyl}
\end{equation}
which is a strong operator identity by the very nature of its derivation.

\subsection{Unlike spin coherent-state closure relation}
To show that our analysis is not limited to canonical coherent states, we now address spin coherent states. 
Let $\hat{\bf J}=(\hat{J}_1,\hat{J}_2,\hat{J}_3)$ be a general angular momentum operator in quantum mechanics, i.e., $[\hat{J}_1,\hat{J}_2]=i \hbar \hat{J}_3$, etc. An arbitrary rotation on a ket in a $(2j+1)$-dimensional Hilbert space ($j=0,1/2,1,2/3,2,...$) can be characterized by the operator
$\hat{W} \propto (w\tilde{J}_--w^*\tilde{J}_+-i \varphi\tilde{J}_3)$, $w$ being an arbitrary complex number and $\varphi$ an angle. Also $\tilde{\bf J}\equiv \hat{\bf J}/\hbar$, $\hat{ J}_{\pm}=(\hat{ J}_1\pm i\hat{J}_2)/\sqrt{2}$. A spin coherent state is defined by $| w \rangle={\cal N}\exp\{\hat{W} \}|y \rangle$, where ${\cal N}$ stands for a normalization constant~\cite{klauder2} and $| y \rangle$ is an arbitrary reference state in the $(2j+1)$-dimensional Hilbert space. If we choose this state to be an eigenstate of $\tilde{J}_3$, and more specifically, the one with the larger eigenvalue, $\tilde{J}_3| y\rangle=j|y\rangle=j|j\rangle$, then we simply get $| w \rangle={\cal N}\exp\{w\tilde{J}_- \}|j \rangle$, with ${\cal N}=(1+|w|^2)^{-j}$. In this context it is easy to show that the inner product is 
\begin{equation}
\langle w|w'\rangle=(1+|w|^2)^{-j}(1+|w'|^2)^{-j}(1+w^*w')^{2j}\;,
\end{equation} 
and that the resolution of unity can be written in terms of continuous complex variables as
\begin{equation}
\hat{I} = \frac{2j+1}{\pi}\int \frac{{\rm d}^2 w}{(1+|w|^2)^2} \;| w \rangle \langle w| =\sum_{n=-j}^{j} |n\rangle \langle n|\;,
\label{IS}
\end{equation}
with $\tilde{J}_3|n\rangle=n|n\rangle$.

Our natural candidate for a spin coherent state unlike closure relation is 
\begin{equation}
\hat{I} = \frac{\lambda(2j+1)}{\pi}\int \frac{{\rm d}^2 w}{(1+\lambda|w|^2)^2} \;\frac{|\lambda w \rangle \langle w|}{ \langle w|\lambda w \rangle}\;,
\label{IS2}
\end{equation}
which is analogous to identity (\ref{I1}) with $\lambda$ also being a real, positive number. The demonstration is quite simple. Note that
\begin{equation}
\frac{|\lambda w \rangle \langle w|}{ \langle w|\lambda w \rangle}=(1+\lambda r^2)^{-2j}\sum_{n,m}\frac{\lambda^m}{n!m!}r^{n+m}e^{i\phi(n-m)}\tilde{J}_-^n|j\rangle \langle j|\tilde{J}_+^m\;,
\end{equation}
where we employed polar variables $w=re^{i \phi}$. The angular integration gives $2\pi \delta_{n,m}$, and proceeding the change $x=\lambda r^2$, it is immediate that (\ref{IS2}) reduces to $\sum_{n=0}^{2j} |j-n\rangle \langle j-n|=\sum_{n=-j}^{j} |n\rangle \langle n|$, showing that (\ref{IS2}) is indeed a closure relation.
\section{Path integrals and weak energy values}
In this section we apply identity (\ref{Iweyl}) to build a phase-space path integral in which the role of the Hamiltonian is played by a 
weak energy value. Below we give the analogous of Klauder's first form of the path integral \cite{klauder2}. We intend to evaluate the propagator
$K(z',z'',T)=\langle z''|\exp\{ -iT\hat{H}/\hbar\}|z'\rangle$,
where $\hat{H}$ is the Hamiltonian operator. Defining $ z''\equiv z_{N+1}+\zeta_{N+1}$, $z'\equiv z_{0}-\zeta_{0}$, and $\tau\equiv T/(N+1)$ one can write
\begin{equation}
K(z',z'',T)=\lim_{N\rightarrow \infty}\langle z_{N+1}+\zeta_{N+1}| \left( \hat{I}-\frac{i\tau\hat{H}}{\hbar} \right)^{N+1}| z_{0}-\zeta_{0} \rangle\;.
\end{equation}
The limit $N \rightarrow \infty$ is taken along with $\tau \rightarrow 0$ such that the product $\tau (N+1)=T$ remains constant. By inserting $N$ unit operators (\ref{Iweyl}) between the products we obtain
\begin{eqnarray}
\nonumber
K(z',z'',T)=\lim_{N\rightarrow \infty}\int \frac{{\rm d}^2 z_1}{\pi} \dots \frac{{\rm d}^2 z_N}{\pi}\; \frac{\langle z_{N+1}+\zeta_{N+1}| ( \hat{I}-i\tau\hat{H}/\hbar )| z_{N}-\zeta_{N} \rangle}{\langle z_N+\zeta_N|z_N-\zeta_N\rangle}\\
\nonumber
\dots \frac{\langle z_{j+1}+\zeta_{j+1}| ( \hat{I}-i\tau\hat{H}/\hbar )| z_{j}-\zeta_{j} \rangle}{\langle z_j+\zeta_j|z_j-\zeta_j\rangle} \dots\\
\frac{\langle z_{2}+\zeta_{2}| ( \hat{I}-i\tau\hat{H}/\hbar )| z_{1}-\zeta_{1} \rangle}{\langle z_1+\zeta_1|z_1-\zeta_1\rangle}\; 
\langle z_{1}+\zeta_{1}| ( \hat{I}-i\tau\hat{H}/\hbar)| z_{0}-\zeta_{0} \rangle\;.
\end{eqnarray}
We, thus, get an inconvenient asymmetry, since there is no $\langle z_0+\zeta_0|z_0-\zeta_0\rangle$ in the denominator of the last term. This difficulty can be circumvented at the cost of an extra constraint, namely, $\zeta_{0}=0$, implying $z_0=z'$. In the continuum $\zeta$ becomes a function of time, so that the previous condition reads $\zeta(0)=0$. With this, it is harmless to write
\begin{equation}
K(z',z'',T)=\lim_{N\rightarrow \infty}\int \prod_{n=0}^{N} \frac{\langle z_{n+1}+\zeta_{n+1}| (\hat{I}-i\tau\hat{H}/\hbar )| z_{n}-\zeta_{n} \rangle}{\langle z_n+\zeta_n|z_n-\zeta_n\rangle}\prod_{n=1}^N\frac{{\rm d}^2 z_n}{\pi}\;.
\end{equation}
Let us recast the numerator in this expression as 
\begin{eqnarray}
\nonumber
\langle z_{n+1}+\zeta_{n+1}| z_{n}-\zeta_{n} \rangle-i\tau/\hbar\langle z_{n+1}+\zeta_{n+1}|\hat{H}| z_{n}-\zeta_{n} \rangle\\
\nonumber
= \langle z_{n+1}+\zeta_{n+1}| z_{n}-\zeta_{n} \rangle \left(1-\frac{i\tau}{\hbar}{\cal H}_{n}\right)\\
=\langle z_{n+1}+\zeta_{n+1}| z_{n}-\zeta_{n} \rangle \exp\left\{ -\frac{i\tau}{\hbar}{\cal H}_{n}\right\}\;,
\end{eqnarray}
where 
\begin{equation}
{\cal H}_{n}=\frac{\langle z_{n+1}+\zeta_{n+1}|\hat{H}| z_{n}-\zeta_{n} \rangle}{\langle z_{n+1}+\zeta_{n+1}| z_{n}-\zeta_{n} \rangle}\;.
\label{hn}
\end{equation}
Therefore
\begin{eqnarray}
K(z',z'',T)=\lim_{N\rightarrow \infty}\int \prod_{n=0}^{N} F_n 
\;\exp\left\{ -\frac{i\tau}{\hbar}{\cal H}_{n}\right\}\prod_{n=1}^N\frac{{\rm d}^2 z_n}{\pi}\;,
\label{path1}
\end{eqnarray}
with
\begin{equation}
F_n=\frac{\langle z_{n+1}+\zeta_{n+1}| z_{n}-\zeta_{n} \rangle}{\langle z_n+\zeta_n|z_n-\zeta_n\rangle} \;.
\end{equation}
Expression (\ref{path1}) represents a valid discrete version of a phase-space path integral. Typically, the paths that enter in the evaluation of (\ref{path1}) are nowhere continuous. However, it is helpful, although not rigorously justifiable, to imagine the paths to be continuous and differentiable and take the limit $N\rightarrow \infty$ before proceeding to the integrations. This assumption becomes more reasonable in the semiclassical regime, since in this limit we expect that the contributing paths are in the vicinity of the classical (smooth) trajectory. The key point is that, in this case, one can write
$z_{n+1}+\zeta_{n+1}\equiv z_{n}+\zeta_{n}+\delta z_{n}+\delta \zeta_{n}$, where $|\delta z_{n}+\delta \zeta_{n}|\rightarrow 0$
for $\tau \rightarrow 0$. To first order in $\delta z_{n}$ and $\delta \zeta_{n}$ we get
\begin{eqnarray}
\nonumber
F_n=\exp\left\{ -\frac{1}{2}(\delta z_{n}+\delta \zeta_{n})^*(z_{n}+\zeta_{n}) \right.\\
\left.-\frac{1}{2}(\delta z_{n}+\delta \zeta_{n})(z_{n}+\zeta_{n})^*+(\delta z_{n}+\delta \zeta_{n})^*( z_{n}-\zeta_{n}) \right\}\;.
\end{eqnarray}
Exchanging the ordering of integrations and products and taking the limit $N \rightarrow \infty$, we get 
\begin{equation}
K(z',z'',T)=\int\exp\left\{ \int_0^T {\rm d}t\,F(t) -\frac{i}{\hbar}\int_0^T {\rm d}t\,{\cal H}_{\zeta} \right\}\,{\cal D}z\;,
\end{equation}
where ${\cal D}z\equiv \lim_{N \rightarrow \infty}\prod_{n=1}^N\frac{{\rm d}^2 z_n}{\pi}$, and
\begin{equation}
\label{hfraco}
{\cal H}_{\zeta}=\frac{\langle z+\zeta|\hat{H}| z-\zeta \rangle}{\langle z+\zeta| z-\zeta \rangle}\;
\end{equation}
is the continuous counterpart of (\ref{hn}). The discrete quantity $F_n$ becomes
\begin{eqnarray}
F(t)=-\frac{\rm d}{{\rm d}t}[\zeta(z^*+\zeta^*)]+\frac{1}{2}(z-\zeta)\frac{\rm d}{{\rm d}t}(z+\zeta)^*
-\frac{1}{2}(z+\zeta)^*\frac{\rm d}{{\rm d}t}(z-\zeta)\;.
\end{eqnarray}
Finally one can write the formal expression for the path integral as
\begin{equation}
\label{path2}
K(z',z'',T)=\int \exp\left\{ -\zeta(T)[{z''}^*+\zeta^*(T)] +\frac{i}{\hbar}S_{\zeta} \right\}\,{\cal D}z\;,
\end{equation}
where the first expression in the argument of the exponential is a surface term for which we already employed the condition $\zeta(0)=0$. The last term is a generalized action
\begin{equation}
S_{\zeta} =\int_0^T\,\left[\frac{i\hbar}{2}(z+\zeta)^*(\dot{z}-\dot{\zeta}) \\
-\frac{i\hbar}{2}(z-\zeta)(\dot{z}+\dot{\zeta})^*-{\cal H}_{\zeta} \right]\,{\rm d}t \;,
\label{Sfraca}
\end{equation}
where the dot denotes time derivative.
For $\zeta\equiv 0$ the surface term vanishes and we get $S_0=\int_0^T\,{\rm d}t [i\hbar(z^*\dot{z}-z\dot{z}^*)/2-{\cal H}_0]=\int_0^T\,{\rm d}t [(p\dot{q}-q\dot{p})/2-{\cal H}_0]$, ${\cal H}_0=\langle z|\hat{H}|z\rangle$, as expected. While ${\cal H}_0=\langle z| \hat{H} | z \rangle$ is a real function of the phase-space coordinates $q$ and $p$, ${\cal H}_{\zeta}$ is, in general, complex valued. This might seem a strong disadvantage of expression (\ref{path2}), but, in fact, it is not. The functions ${\cal H}_{0}$ and ${\cal H}_{\zeta}$ fully assume the role of Hamiltonians into classical equations of motion only in the semiclassical limit. It is well known, however, that in this regime, even for ${\cal H}_{0}$, the classical trajectories are, so to speak, overloaded with boundary conditions [$z^*(0)={z'}^*$ and $z{(T)}={z''}$], which can be satisfied only by extending both $q(t)$ and $p(t)$ to the complex plane. In this context, a complex function as the effective Hamiltonian is fairly natural.

It is worth to note that (\ref{hfraco}) is a weak energy value as originally defined by Aharonov, Albert, and Vaidman \cite{aharonov}, see Eq. (\ref{weak})
for $\langle \psi_f| \psi_0 \rangle \ne 0$ (which is always fulfilled in the present case). In fact, in addition, we should have $\langle \psi_f| \psi_0 \rangle \approx 1$, which would demand $|\zeta|<<1$ in (\ref{hfraco}). Thus, in this regime, the complex number ${\cal H}_{\zeta}$, is a weak value of energy related to the states $| \psi_0 \rangle =| z-\zeta \rangle$ and $|\psi_f\rangle=| z+\zeta \rangle$. This kind of weak value has been studied in \cite{lars} and shown to make the nonclassical properties of coherent states explicit. For small $|\zeta|$ one can write
\begin{eqnarray}
\nonumber
{\cal H}_{\zeta}=\frac{\langle z|\hat{D}^{\dagger}(\zeta)\hat{H}\hat{D}(-\zeta)| z \rangle}{\langle z|\hat{D}^{\dagger}(\zeta)\hat{D}(-\zeta)| z \rangle}\approx
\frac{{\cal H}_0-\zeta\langle z| \{\hat{H},\hat{a}^{\dagger}\}| z \rangle+\zeta^*\langle z| \{\hat{H},\hat{a}\}| z \rangle }{1-2\zeta z^*+2\zeta^*z}\\
\approx {\cal H}_0-\zeta\langle z| \{\hat{H},\hat{a}^{\dagger}\}| z \rangle+\zeta^*\langle z| \{\hat{H},\hat{a}\}| z \rangle +2(\zeta z^*-\zeta^* z){\cal H}_0\;,
\end{eqnarray}
where $\{\, ,\,\}$ stands for the anticommutator, $ {\cal H}_0=\langle z|\hat{H}| z \rangle$ and we used $\hat{D}(-\zeta)\approx \hat{I}-\zeta\hat{a}^{\dagger}+\zeta^*\hat{a}$. After reordering operators, the previous expression can be written as 
\begin{equation}
{\cal H}_{\zeta}={\cal H}_0+\zeta^*\langle z| [\hat{a},\hat{H}]| z \rangle+\zeta\langle z| [\hat{a}^{\dagger},\hat{H}]| z \rangle+O(|\zeta|^2)\;.
\end{equation}
Here we must be careful in handling the expectation values by noting that $\langle z(t)| [\hat{a},\hat{H}]| z(t) \rangle \ne i \hbar \dot{z}(t)$, since $| z(t) \rangle$ is not, in general, a solution of the time-dependent Schr\"odinger equation.
\subsection{Weak values in the quasi-classical domain}
Nonetheless, when the system scales (size, energy, etc) are such that $S_{class}>>\hbar$, given that $|\psi(0)\rangle$ is a coherent state, it will remain so for a time which is longer for 
larger ratios $S_{class}/\hbar$, with the center of the packet following, to first order, the classical trajectory. In this quasi-classical regime $|\psi(t)\rangle\approx|z_{class}(t)\rangle$. It is also a well known result that ${\cal H}_0=H_{class}+O(\hbar)$ \cite{klauder,marcus}.
In this limit one can write the quasi-classical weak energy value as
\begin{equation}
{\cal H}_{\zeta}\approx H_{class}+i \hbar (\zeta^*\dot{z}_{class}+\zeta \dot{z}^*_{class})=
H_{class}+i \left[\alpha X \,\dot{q}_{class}+\frac{\Pi \,\dot{p}_{class}}{\alpha}\right]\;.
\end{equation}
with $\zeta \equiv X/\sqrt{2}b+ib\Pi/\sqrt{2}\hbar$ and recalling that $b=\sqrt{\hbar/\alpha}$, where $\alpha$ is a constant with dimension of $mass/time$ (for the harmonic oscillator $\alpha=m\omega$). Therefore, the real part of the quasi-classical weak energy is
the Hamiltonian itself, while the imaginary part (first order in $|\zeta|$ and zeroth order in $\hbar$) depends on the tangent field in phase space. By replacing this into (\ref{Sfraca}) we get, with a slightly abusive language, the associated weak action integral.

\section{Discussion and Conclusion}
About ninety years after Schr\"odinger discovered coherent states as quasi-classical minimum uncertainty wave functions \cite{sch, sch2}, they can still offer us some surprise. 
The structure behind overcomplete tight frames is richer if they fulfill the mutual non-orthogonality property  [we say that $\{|v\rangle\}$ is a mutually overlapping (or mutually non-orthogonal) frame if $\langle v|v'\rangle\ne 0$ for all possible pairs $v$ and $v'$].
Under this condition we showed that all known unlike closure relations can be expressed as:
\begin{equation}
\label{conj}
\hat{I} \propto \int {\rm d}V \frac{| f(v) \rangle \langle v|}{\langle v| f(v)\rangle}\;,
\end{equation} 
where d$V \propto {\rm d} \Re(v) {\rm d} \Im(v)$ is a volume element (d$\mu=\langle v| f(v)\rangle^{-1}$ d$V$). The fact that $| v \rangle$ has a non-zero projection onto every $|f(v)\rangle=| v' \rangle$ in the set,
enlarges the notion of component or coefficient of a vector, such that $\langle v|\psi \rangle$ can be made the component associated to $| v' \rangle$, provided that this is accompanied by a suitable correction by an {\it amplifying} measure, $\langle v| v'\rangle^{-1}$, where the amplification $1\le|\langle v| v'\rangle|^{-1} < \infty$ is larger for larger Euclidian distances $|| v \rangle-| v' \rangle|$.

It is curious to realize that the appearance of this kind of weighting factors in quantum mechanical integrals is not unusual, although, some times concealed. As an example consider the Weyl symbol of an arbitrary operator $\hat{A}$ \cite{alfredo}, given by 
\begin{equation}
A_W=\int {\rm d} x \, \langle q+x/2|\hat{A}|q-x/2\rangle e^{-ipx/\hbar}\;,
\end{equation}
and note that it can be expressed as
\begin{equation}
A_W=\frac{1}{2\pi \hbar}\int {\rm d} x \, \frac{\langle q+x/2|\hat{A}|q-x/2\rangle}{\langle q+x/2|p\rangle \langle p|q-x/2\rangle} \;.
\end{equation}
Another case, belonging to a slightly different category, is the recently defined dual representation to the Bargmann function $\psi(z^*)$ \cite{marcus2, dual}, that has found some application in semiclassical physics \cite{marcus2, alexandre} and in quantum gravity \cite{magueijo,frassek}. It is given by
\begin{equation}
\label{dual}
f_{\psi}(w)=\int_{\gamma}{\rm d}z^*\, \psi(z^*)\,e^{-z^*w}=\int_{\gamma}{\rm d}z^*\,\frac{\langle z| \psi \rangle}{\langle z| w \rangle}\;,
\end{equation}
where $\gamma$ is a curve in the complex plane. The previous definition can be seen as an extension of a Fourier transform connecting the wave function $\psi(x)=\langle x| \psi \rangle$ to the momentum representation
\begin{equation}
\label{four}
\tilde{\psi}(p)=\frac{1}{\sqrt{2\pi \hbar}}\int \, {\rm d}x\, \psi(x) e^{-\frac{i}{\hbar}px}=\frac{1}{2\pi \hbar}\int \, {\rm d}x \,\frac{\langle x| \psi \rangle}{\langle x| p \rangle}\;.
\end{equation}
These transformations can be mnemonically seen as though the two bra's $\langle z|$ in (\ref{dual}) and $\langle x|$ in (\ref{four}) take part in a sort of cancelation.

The question raised about operators of the form (\ref{conj}), or more specifically, on which further functions $f(v)$, if any, would lead to resolutions of unity, seems to be one worth of some thought. The failure of the simple unilike resolution (\ref{Idiscrete}) in the appendix to have a valid counterpart in the Hilbert space seems to indicate that analyticity is a necessary ingredient, that is, $\exp\{|v|^2/2\}\langle \psi|f(v) \rangle$  should be analytic functions of $v$ and $v^*$, respectively.

As for the alternative form of the coherent-state path integral we presented here, it is hoped that it may be useful, e. g., in attenuating root-search problems in the semiclassical dynamics \cite{adachi,kay}. A completely analogous procedure can be adopted to derive yet another form of the path integral starting from (\ref{I1}). However, this does not seem to bring any relevant new information, unless a specific application arises.

The issues addressed in this work are related to the early work \cite{parisio} and to the ${\cal D}$-pseudo-boson formalism developed more recently \cite{dpb}, in particular, to the examples given in \cite{ali} [see eq. (2.4) in this reference].

Finally, it is also interesting to note that we reached the concept of weak values with no reference to the delicate concept of quantum measurement.

\appendix

\section{Unlike closure relations in $\mathbb{R}^2$}

In this appendix we illustrate some aspects of coherent-state overcompleteness with a toy construction in the Euclidean plane. 
A similar example can be found in \cite{gazeau}, but here we go further to ensure the mutual overlapping property. 

Consider an arbitrary orthonormal basis $\{|U \rangle, |V \rangle \}$ and consider the overcomplete basis composed of $N$ normalized vectors given by 
$|Z_n \rangle \equiv\cos\left(\frac{n \Delta \theta}{N}\right)|U\rangle+\sin \left(\frac{n \Delta \theta}{N}\right)|V\rangle$,
with $n=1,2,...,N$ and define the operator $\hat{A}_N=\frac{2}{N}\sum_{n=1}^{N} |Z_n\rangle \langle Z_n|$.

We initially assume that $\Delta \theta=2\pi$, so that, the $N$ vectors have directions uniformly distributed over $(0,2\pi]$, with step $2\pi/N$. Taking the 
continuum limit $\frac{1}{N}\sum_{n=1}^{N} \rightarrow \frac{1}{2\pi}\int_0^{2\pi} {\rm d} \theta$, we get the well known result
\begin{equation}
\label{A}
\hat{A}_{\infty}=\lim_{N\rightarrow \infty} \frac{2}{N}\sum_{n=1}^{N} |Z_n\rangle \langle Z_n|=|U\rangle \langle U|+ |V\rangle \langle V|=\hat{I}\;,
\end{equation}
which resembles the quantum coherent state relation (\ref{I0}) in its form, and in the sense that a redundant set of vectors is employed to represent the resolution of unity with uniform weight. Here the usual feature of overcompleteness appears: any pair of non-degenerate vectors $\{| Z_r\rangle, |Z_s\rangle\}$ suffices to generate arbitrary vectors in $\mathbb{R}^2$. This fact is the direct analogous of quantum representations using subsets of the $z$-plane \cite{circle,line}. 
The above lines essentially correspond to the construction given by Gazeau in \cite{gazeau}.

\begin{figure}
\begin{center}
\includegraphics[width=6cm,angle=0]{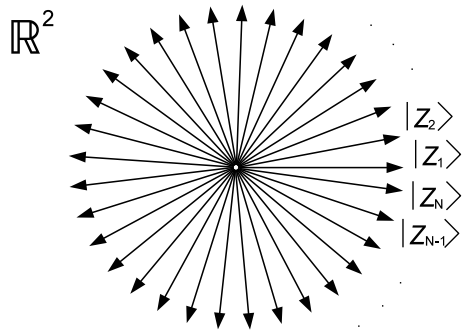}
\end{center}
\caption{Set $\{ |Z_n \rangle\}$ of $N=33$ mutually non-orthogonal unit vectors separated by a constant angle $\Delta \theta=(2-\epsilon)\pi /(N-1)$, $\epsilon=\sqrt{2}/35 \approx 0.04$. The angle between $|Z_N \rangle$ and $|Z_1 \rangle$ is not $\Delta \theta$. This asymmetry becomes less relevant for increasing values of $N$.}
\label{figure1}
\end{figure}

For our purposes, however, the previous analogy is still insufficient because it disregards the essential fact that $\langle z'|z \rangle \ne 0$. In the present case $\langle Z_n|Z_m\rangle$ does vanish if $(n-m)/N=1/4, 3/4$. This difficulty can be avoided if we assume
$\Delta \theta=(2-\epsilon)\pi$, where $\epsilon$ is an irrational number that can be made {\it arbitrarily small} from the outset. The condition of orthogonality reads $2-\epsilon=N/[2(n-m)]$ or $2-\epsilon=3N/[2(n-m)]$. In both cases we have an irrational number in the left-hand side and a rational number in the right-hand side, thus, ensuring that any pair of vectors in the frame is non-orthogonal. 
The price to be paid is that there will be a residual anisotropy in the set $\{ |Z_n \rangle \}$ [see figure \ref{figure1}]. With this, the operator $\hat{A}_{\infty}$, (\ref{A}), becomes $\hat{I}+O(\epsilon)$. This apparently futile detail is important due to the nature of the unusual closure relation we deal with in what follows. 

Specifically, let us investigate if it is possible to express the resolution of unity in terms of single sums of unlike outer products $|Z_k\rangle \langle Z_n|$, $k=k(n)\ne n$. We intend to write $\hat{I} \propto\sum_n \mu (n) |Z_{k(n)}\rangle \langle Z_n|$, were $\mu (n)$ is a correction due to the projection of the $n$th component in the distinct direction $| Z_{k(n)}\rangle$. The point we want to stress is that 
\begin{equation}
\mu (n)=\langle Z_n|Z_{k(n)}\rangle^{-1}\;,
\end{equation}
which is well defined because of the mutual overlapping property, does the job. As an example take $k(n)=N-n$ and define the operator
\begin{equation}
\hat{B}_N=\frac{2}{N}\sum_{n=1}^{N} \frac{|Z_{N-n}\rangle \langle Z_n|}{\langle Z_n|Z_{N-n}\rangle}=\frac{2}{N}\sum_{n=1}^{N} \frac{|Z_{N-n}\rangle \langle Z_n|}{\cos[\Delta \theta(1-2n/N)]}\;.
\end{equation} 
It can be easily shown that in the limit $N\rightarrow \infty$ we get
\begin{equation}
\hat{B}_{\infty}= (1+L)|U\rangle \langle U|+(1-L)|V\rangle \langle V|+J_+|U\rangle \langle V|+J_-|V\rangle \langle U|\;,
\end{equation}
where
\begin{eqnarray}
L=\frac{\cos \Delta \theta}{\Delta \theta}\int_{0}^{\Delta \theta}\sec (2\theta-\Delta \theta) \,{\rm d}\theta = O(\epsilon)\;,\\
J_{\pm}=\frac{\tan \Delta \theta }{\Delta \theta}\,L\pm \frac{1 }{\Delta \theta}\int_{0}^{\Delta \theta} \tan (2\theta-\Delta \theta) \,{\rm d}\theta=O(\epsilon^2)\;.
\end{eqnarray}
This leads to
\begin{equation}
\hat{I}=\lim_{N\rightarrow \infty}\frac{2}{N}\sum_{n=1}^{N} \frac{|Z_{N-n}\rangle \langle Z_n|}{\langle Z_n|Z_{N-n}\rangle}+O(\epsilon)\;,
\label{Idiscrete}
\end{equation}
where we used $\Delta \theta =(2-\epsilon)\pi$, with $\epsilon$ being an arbitrary irrational that can be made as small as needed from the beginning.
Note that this kind of closure relation is ill defined if one deals with an orthonormal basis, since all terms $\langle e_i|e_j \rangle^{-1}$ would diverge for $i \ne j$.
\label{ap}

\end{document}